\documentclass[conference]{IEEEtran}
\IEEEoverridecommandlockouts
\usepackage{cite}
\usepackage{amsmath,amssymb,amsfonts}
\usepackage{algorithmic}
\usepackage{graphicx}
\usepackage{textcomp}
\usepackage{xcolor}
\def\BibTeX{{\rm B\kern-.05em{\sc i\kern-.025em b}\kern-.08em
    T\kern-.1667em\lower.7ex\hbox{E}\kern-.125emX}}
\begin{document}

\title{Decoding Visual Imagery from EEG Signals using Visual Perception Guided Network Training Method\\

\thanks{This work was partly supported by Institute of Information \& Communications Technology Planning \& Evaluation (IITP) grant funded by the Korea government (MSIT) (No. 2017-0-00432, Development of Non-Invasive Integrated BCI SW Platform to Control Home Appliances and External Devices by User’s Thought via AR/VR Interface; No. 2017-0-00451, Development of BCI based Brain and Cognitive Computing Technology for Recognizing User’s Intentions using Deep Learning; No. 2019-0-00079, Artificial Intelligence Graduate School Program, Korea University).}
}

\author{\IEEEauthorblockN{Byoung-Hee Kwon}
\IEEEauthorblockA{\textit{Dept. Brain and Cognitive Engineering} \\
\textit{Korea University}\\
Seoul, Korea \\
bh\_kwon@korea.ac.kr}

\and

\IEEEauthorblockN{Jeong-Hyun Cho}
\IEEEauthorblockA{\textit{Dept. Brain and Cognitive Engineering} \\
\textit{Korea University}\\
Seoul, Korea \\
jh\_cho@korea.ac.kr}

\and

\IEEEauthorblockN{Byeong-Hoo Lee}
\IEEEauthorblockA{\textit{Dept. Brain and Cognitive Engineering} \\
\textit{Korea University}\\
Seoul, Korea \\
bh\_lee@korea.ac.kr}

}

\maketitle

\begin{abstract}
An electroencephalogram is an effective approach that provides a bidirectional pathway between user and computer in a non-invasive way. In this study, we adopted the visual perception data for training the visual imagery decoding network. We proposed a visual perception-guided network training approach for decoding visual imagery. Visual perception decreases the power of the alpha frequency range of the visual cortex over time when the user performed the task, and visual imagery increases the power of the alpha frequency range of the visual cortex over time as the user performed with the task. Generated brain signals when the user performing visual imagery and visual perception have opposite brain activity tendencies, and we used these characteristics to design the proposed network. When using the proposed method, the average classification performance of visual imagery with the visual perception data was 0.7008. Our results provide the possibility of using the visual perception data as a guide of the visual imagery classification network training.\\
\end{abstract}

\begin{IEEEkeywords}
brain-computer interface, visual imagery, visual perception, deep learning
\end{IEEEkeywords}

\section{Introduction}
Brain-computer interface (BCI) is the technology that creates communication pathways between users and computers possible using brain signals \cite{vaughan2003brain, zhang2019strength, kwon2019subject}. Electroencephalography (EEG) has the advantage of higher time resolution than comparable methods such as functional magnetic resonance imaging\cite{zhang2017hybrid} and near-infrared spectroscopy \cite{chen2016high, lee2017network}. Therefore, in this study, we applied the visual imagery that a type of EEG-based BCI endogenous paradigm.

Recently, various studies have been conducted to decode human intentions using brain signals \cite{jeong2018decoding, he2018brain, lee2020continuous, chholak2019visual, suk2014predicting}. In particular, in order to control BCI-related devices, many related studies were conducted to collect and decode EEG signals related to users' intentions using various BCI paradigms. In order to control BCI-related devices, paradigms such as steady-state visual evoked potential (SSVEP), \cite{won2015effect, kwak2017convolutional}, P300 \cite{yeom2014efficient, lee2018high}, and motor imagery (MI) \cite{kim2014decoding, zhang2021adaptive, kam2013non, jeong2020brain} were used in related studies. Exogenous paradigms such as SSVEP and P300 require external devices and cause disadvantages such as a decrease in user concentration and accumulation of fatigue. Also, MI is perceived differently by each user, therefore it causes a lack of uniformity. As a result, there can be a discrepancy between the user intention and what the user imagines.

To overcome these limitations, in this study we used visual imagery to decode the user's intention. Visual imagery is a paradigm that based on visual perception experience and does not require additional external devices \cite{kwon2020decoding}. When the user performs visual imagery, the user imagines a picture or movement as if drawing a picture. Visual imagery has a wide range of brain signals from the frontal area to the occipital area containing the visual cortex. Specifically, visual imagery can be analyzed in various frequency ranges such as delta, theta, and alpha band, and the prefrontal and occipital lobe are mainly activated \cite{koizumi2018development}. Brain activities based on visual imagery induce delta band in the prefrontal lobe and the alpha band in the occipital lobe \cite{sousa2017pure}. There are clear evidence that the visual imagery and visual perception can be decoded from visual cortex including V1 and V2. 

\begin{figure*}[t!]
\centering
\centerline{\includegraphics[width=0.96\textwidth]{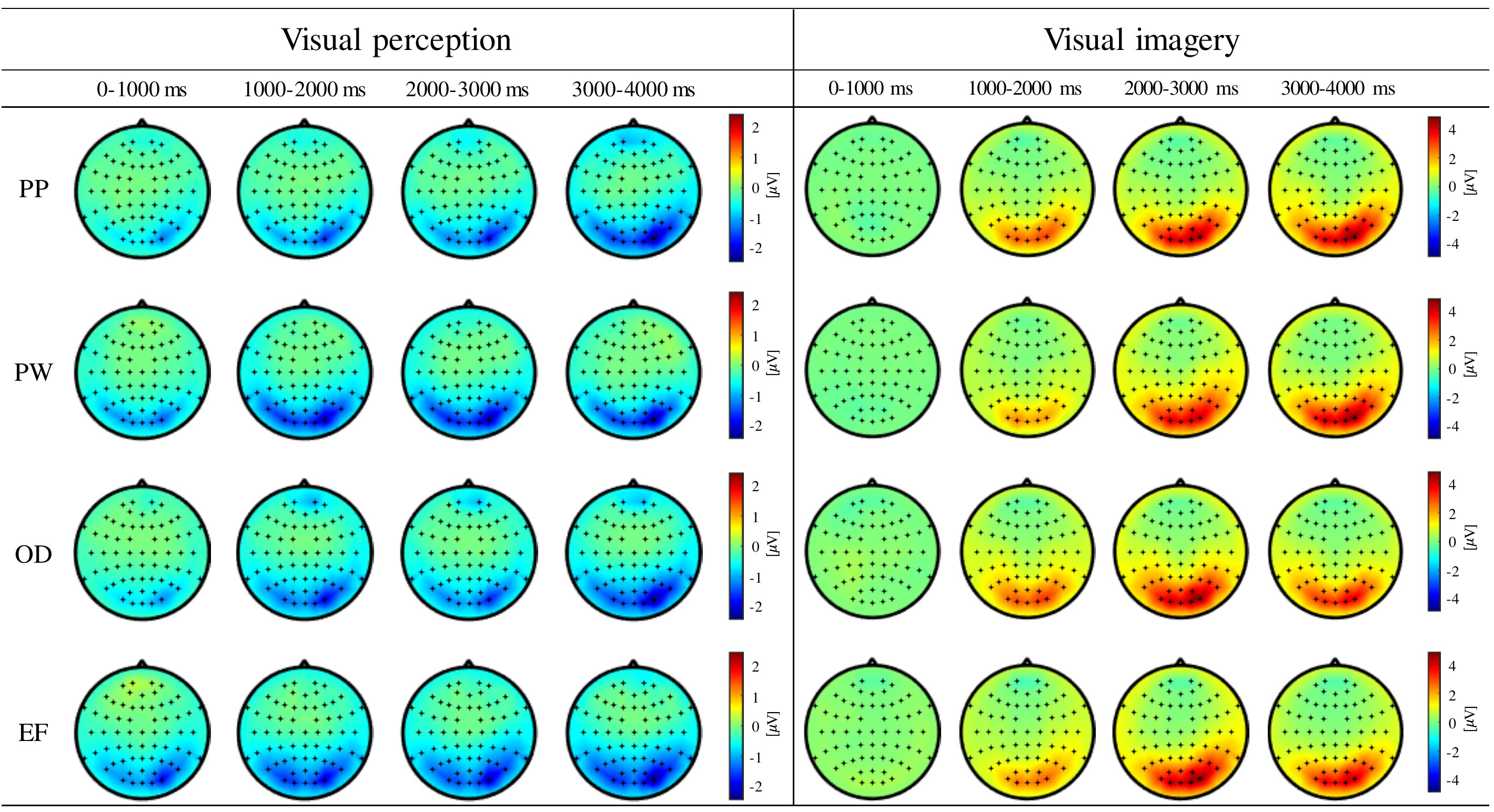}}
\caption{The representation of brain activation in the alpha frequency range for data when S01 performs visual perception and visual imagery in each class. PP, PW, OD, and EF represent picking up a cell phone, pouring water, opening door, and eating food respectively. The activation of visual perception decreases in the occipital lobe in the alpha frequency range. However, the activation of visual imagery increases in the occipital lobe in the alpha frequency range.}
\label{fig:1}
\end{figure*}

Visual perception refers to when a user looks at an object, which manifests a specific brain signal in the visual cortex. Brain signals expressed during visual imagery have a similar direction to visual perception, and their intensity and degree of similarity become stronger over time. Visual perception reduces the brain activity of the alpha frequency range in the visual cortex over time as the user proceeds with the task, and visual imagery increases the brain activity of the alpha frequency range in the visual cortex over time as the user proceeds with the task. In this study, we used the brain activity tendencies of visual perception and visual imagery to reduce the differences between the two types of data and construct a network accordingly to improve the decoding performance of visual imagery. The classes of visual imagery consisted of a total of four tasks (pouring water, opening door, eating food, and picking up a cell phone), and instructed the user to perform a task on the black screen when performing visual imagery in order to clearly give the difference between visual perception and visual imagery. In this study, we confirmed the possibility that the use of both data at the same time could lead to the effect of data augmentation. Also, we can infer that visual perception can guide the training of the visual imagery classification network.

\section{Methods}

\subsection{Dataset} 
We used the EEG data from our previous study \cite{kwon2020decoding}, from eight subjects (S01–S08; ages 24-30 (Mean: 26.6,SD: 1.89); 4 men and 4 women, all right-handed). EEG data were collected by 1,000 Hz sampling rate using 64 Ag/AgCl electrodes (Fp1--2, AF3--4, AF7--8, AFz, F1--8, Fz, FC1--6, FT7--10, C1--6, Cz, T7--8, CP1--6, CPz, TP7--10, P1--8, Pz, PO3--4, PO7--8, POz, O1--2, Oz, Iz) in the 10/20 international system via BrainAmp (BrainProduct GmbH, Germany). 

The data consists of the separation of the visual imagery phase and the visual perception phase to collect the high quality of visual imagery and visual perception-related EEG signals. Also, between the visual imagery phase and visual perception phase, there is a 5-s length rest phase to prevent the after-effect of the previous visual stimulus. Each class consisted of 50 trials and a total of 200 trials for visual imagery and 2000 trials for visual perception were collected for every subject.

\subsection{Data Analysis}
We used BBCI toolbox and openBMI \cite{lee2019eeg} with MATLAB 2020a (MathWorks Inc., USA) for pre-processing. The band-pass filter was conducted between [8--13] Hz. Alpha frequency range is the significant frequency range in visual imagery and visual perception. We calculated the change of the power in the alpha frequency range in each channel and represented it in the scalp. As Fig. 1(a) shows, the alpha band power mainly increases in the occipital lobe when the user conducted visual imagery task. In the contrast, the alpha band power decreased in the occipital lobe when the user conducted visual perception task as shown in Fig. 1(b). From these results, brain signals generated when the user performing visual imaging and visual perception have opposite tendencies, and this property can be used to help network training.

\subsection{Proposed Method}
We calculated the changes of the power in the alpha frequency range of visual perception and visual imagery data to understand the characteristics of each data and to use visual perception data for training the network. Using the tendency of the calculated power, visual perception data was modified for the visual imagery classification network. First, we defined the equations for generating modified data $x_{re}$ using the characteristics of the visual related data:

\begin{equation}
{x}_{norm} = {{x}-{x}_{min} \over {x}_{max} - {x}_{min}}
\end{equation}

\begin{table}[t!]
\small
\caption{Architecture Design of Proposed Method}
\renewcommand{\arraystretch}{1.3}
\resizebox{\columnwidth}{!}{%
\begin{tabular}{ccccc}
\hline
Layer & Type        & Output shape                  & Kernel size & Stride \\ \hline
1     & Convolution & 1 × 20 × N of channels × 1192 & 1 × 60      & 1 × 1  \\
2     & Convolution & 1 × 20 × 1 × 1192             & N of channels × 1      & 1 × 1  \\
3     & Convolution & 1 × 40 × 1 × 1163             & 1 × 30      & 1 × 1  \\
4     & Convolution & 1 × 80 × 1 × 1149             & 1 × 15      & 1 × 1  \\
5     & Dropout (0.5) & -              & -       & -  \\
6     & Max Pooling & 1 × 80 × 1 × 164              & 1 × 7       & 1 × 7  \\
7     & Convolution & 1 × 160 × 1 × 150             & 1 × 15      & 1 × 1  \\
8     & Max Pooling & 1 × 160 × 1 × 30              & 1 × 5       & 1 × 5  \\
9     & Convolution & 1 × 320 × 1 × 16              & 1 × 15      & 1 × 1  \\
10     & Max Pooling & 1 × 320 × 1 × 3               & 1 x 5       & 1 × 5  \\
11    & Flatten     & 1 × 960                       & -            & -       \\
12    & Softmax     & 1 × 4                         & -            & -       \\ \hline
\end{tabular}}
\end{table}

where, $x$ represents the data from visual perception. We normalized the data $x$ because the difference of the scale of the amplitude in each data misleads the network training. According to the normalization of the visual perception data $x$, we also normalized the visual imagery data in the same method. 

\begin{equation}
{x}_{re} = {x}_{O} - {x}_{norm}
\end{equation}

We reversed the scaled data ${x}_{norm}$ using $x_O$, zero matrix, with the expectation of having a similar tendency with the visual imagery data.
We used the modified data $x_{re}$ as input of the convolutional neural network (CNN) that we designed (Table I). The proposed network consists of six convolution layers and includes three max-pooling layers. We also applied the dropout layer to prevent the overfitting layer due to the small number of the training dataset. 

\begin{figure}[t!]
\includegraphics[width=\columnwidth]{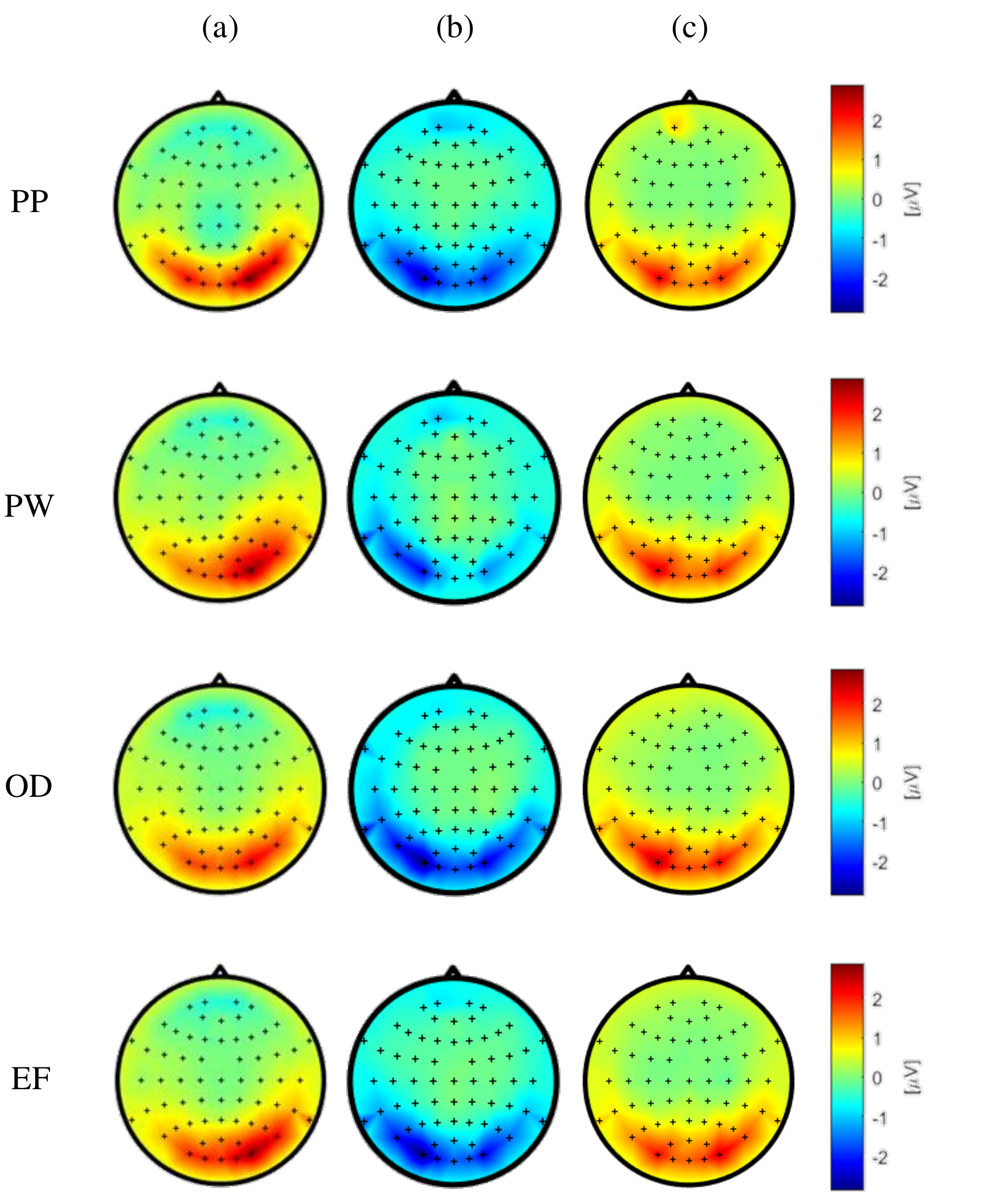}
\caption{The result of power spectral density in visual imagery and visual perception. (a) and (b) indicate the brain activity of the visual imagery and visual perception in the alpha frequency range respectively. (c) indicates the reverse result of visual perception activity.}
\end{figure}

\section{Results and Discussion}
\subsection{Modified Data}
We investigated the changes of power in the alpha frequency range when the user performed visual imagery and visual perception. Fig. 2 shows the representation of the brain activation in the alpha frequency. As Fig. 2(c) shows, the result of modifying the visual perception data to fit the visual imagery data for network training, it was confirmed that the tendency of the modified data was similar to that of the actual imagery data. 

\begin{table*}[t!]
\renewcommand{\arraystretch}{1.3}
\begin{center}
\caption{The classification performance of the proposed method and baseline methods}
\begin{tabular}{cccccccccc}
\hline
                & S01                                                             & S02                                                             & S03                                                             & S04                                                             & S05                                                             & S06                                                             & S07                                                             & S08                                                             & Avg.                                                            \\ \hline
DeepConvNet\cite{schirrmeister2017deep}          & \begin{tabular}[c]{@{}c@{}}0.7205\\ ($\pm$ 0.0213)\end{tabular} & \begin{tabular}[c]{@{}c@{}}0.7230\\ ($\pm$ 0.0683)\end{tabular} & \begin{tabular}[c]{@{}c@{}}0.6375\\ ($\pm$ 0.0885)\end{tabular} & \begin{tabular}[c]{@{}c@{}}0.6480\\ ($\pm$ 0.0208)\end{tabular} & \begin{tabular}[c]{@{}c@{}}0.6680\\ ($\pm$ 0.0537)\end{tabular} & \begin{tabular}[c]{@{}c@{}}0.8272\\ ($\pm$ 0.0621)\end{tabular} & \begin{tabular}[c]{@{}c@{}}0.7005\\ ($\pm$ 0.0556)\end{tabular} & \begin{tabular}[c]{@{}c@{}}0.5540\\ ($\pm$ 0.0131)\end{tabular} & \begin{tabular}[c]{@{}c@{}}0.6848\\ ($\pm$ 0.0480)\end{tabular} \\
EEGNet\cite{lawhern2018eegnet}    & \begin{tabular}[c]{@{}c@{}}0.6520\\ ($\pm$ 0.0242)\end{tabular} & \begin{tabular}[c]{@{}c@{}}0.5880\\ ($\pm$ 0.0568)\end{tabular} & \begin{tabular}[c]{@{}c@{}}0.6945\\ ($\pm$ 0.0789)\end{tabular} & \begin{tabular}[c]{@{}c@{}}0.5512\\ ($\pm$ 0.0554)\end{tabular} & \begin{tabular}[c]{@{}c@{}}0.6030\\ ($\pm$ 0.0403)\end{tabular} & \begin{tabular}[c]{@{}c@{}}0.8215\\ ($\pm$ 0.0953)\end{tabular} & \begin{tabular}[c]{@{}c@{}}0.6910\\ ($\pm$ 0.0921)\end{tabular} & \begin{tabular}[c]{@{}c@{}}0.5420\\ ($\pm$ 0.0852)\end{tabular} & \begin{tabular}[c]{@{}c@{}}0.6429\\ ($\pm$ 0.0660)\end{tabular} \\
Proposed method & \begin{tabular}[c]{@{}c@{}}0.7312\\ ($\pm$ 0.0545)\end{tabular} & \begin{tabular}[c]{@{}c@{}}0.6775\\ ($\pm$ 0.0903)\end{tabular} & \begin{tabular}[c]{@{}c@{}}0.7200\\ ($\pm$ 0.0439)\end{tabular} & \begin{tabular}[c]{@{}c@{}}0.6640\\ ($\pm$ 0.0131)\end{tabular} & \begin{tabular}[c]{@{}c@{}}0.6280\\ ($\pm$ 0.0490)\end{tabular} & \begin{tabular}[c]{@{}c@{}}0.8815\\ ($\pm$ 0.0356)\end{tabular} & \begin{tabular}[c]{@{}c@{}}0.7262\\ ($\pm$ 0.0554)\end{tabular} & \begin{tabular}[c]{@{}c@{}}0.5780\\ ($\pm$ 0.0624)\end{tabular} & \begin{tabular}[c]{@{}c@{}}0.7008\\ ($\pm$ 0.0560)\end{tabular} \\ \hline
\end{tabular}
\end{center}
\end{table*}

\subsection{Performance Evaluation}
We evaluated the proposed method with the baseline methods to verify the possibility of using the modified visual perception data. We used DeepConvNet \cite{schirrmeister2017deep} and EEGNet \cite{lawhern2018eegnet} as baseline methods. In this study, we investigated the classification performance into two perspectives: visual imagery classification with the visual perception data and visual imagery classification without the visual perception data. These results imply that visual perception data can guide the training of the proposed network for visual imagery classification. 

Table II shows the performance comparison using the baseline methods and the proposed method. Using the proposed network, the visual imagery classification performance using visual perception data was $0.7008 (\pm 0.0560)$, which was 0.0160 higher than DeepConvNet and 0.0579 higher than EEGNet. The network we proposed has more convolution layers than the baseline method and is configured to extract local features as the layer deepens to reflect the tendency of visual perception data.

In Table III, classification performances when visual perception data is used and when not used is compared. As a result, the decoding performance of the proposed method was recorded 0.0289 higher when visual perception data was used than when visual imagery data was used only. In addition, when only visual imagery data was used, the proposed network showed classification performance similar to or lower than the baseline methods. This result proves that the network we proposed is an optimized network for using visual perception data.

\begin{table}[t!]
\caption{Performance Comparison using \\ the Conventional Approach and Proposed Approach}
\resizebox{\columnwidth}{!}{
\small
\renewcommand{\arraystretch}{1.7}
\begin{tabular}{ccc}
\hline
                & Visual imagery & Visual imagery + Visual perception \\ \hline
DeepConvNet\cite{schirrmeister2017deep}          & 0.6552         & 0.6848                             \\
EEGNet\cite{lawhern2018eegnet}     & 0.6348         & 0.6429                             \\
Proposed method & 0.6719         & 0.7008                             \\ \hline
\end{tabular}
}
\end{table}

\section{Conclusions and Future Works}
In this study, we modified the visual perception data to guide the training of the visual imagery decoding network that we proposed. Since visual perception and visual imagery have opposite brain activity tendencies in the alpha frequency range, visual perception data must be reversed to decode the visual imagery data. In addition, since brain signals when the user performing visual perception have a larger scale of spectral power than brain signals when the user performing visual imagery, we made similar levels of data by normalizing the two data for network learning in this study. We designed the network to extract local features better as the layer of the network deepens correspond to these data characteristics. Accordingly, when visual perception data was used, the proposed method recorded higher decoding performance than the baseline methods.

However, the proposed network is an approach that forces visual perception data to be similar to visual imagery data by simply using the reverse and normalization approach. In future work, we will develop and apply more advanced approaches to increase the brain activity similarity between the two data using the characteristics of the data.

\bibliographystyle{IEEEtran}
\bibliography{ref}

\end{document}